# On the proper time of the earthquake source

A.V. Guglielmi, O.D. Zotov

*Institute of Physics of the Earth RAS, Moscow, Russia*
*guglielmi@mail.ru,  ozotov@inbox.ru*

**Abstract**

The concept of proper time, which is different from universal time, has been introduced into the physics of earthquakes. Earlier, the idea of proper time was used, but implicitly, and only in a narrow area related to the physics of aftershocks. In this paper, the area of applicability of the concept of proper time is extended, a proper time measurement method is indicated, and an example is given that illustrates the procedure for ordering a sequence of earthquakes in proper time. The global activity of strong earthquakes (M ≥ 7) was chosen as the object of study. We consider the sequence of earthquakes as a random process of the Poisson type. Comparatively weak earthquakes are used as the "underground clock", the ticking of which marks the course of proper time. The Poisson distribution is compared with the distributions for two sequences of strong earthquakes. One of the sequences is ordered by calendar time, and the second by proper time. The result of the test showed that the distribution of events ordered by proper time is closer to the Poisson distribution than the distribution of events ordered by calendar time. We explain this by non-stationarity, which is an immanent property of the Earth's lithosphere.

*Keywords*: geodynamics, aftershocks, Omori's law, evolution equation, deactivation coefficient, non-uniformity of time flow, nonstationarity of geological medium, Poisson process.

## 1. Introduction: "Cooling" source

The idea of proper time of the source arose in the study of aftershocks of a strong earthquake [1, 2]. The idea turned out to be successful and was used in a number of publications (see, for example, [3–7]). The history of question is as follows.

In 1894, twenty-six-year-old Fusakichi Omori discovered the first law of earthquake physics [8], which states that the aftershock frequency $n$, on average, decreases hyperbolically with time:

$$n(t) = \frac{k}{c+t}. \qquad (1)$$

Here $k > 0$, $c > 0$, $t \geq 0$. It was noted [1] that the Omori law (1) can be represented as the aftershock evolution equation

$$\frac{dn}{dt}+\sigma n^2 =0. \qquad (2)$$

Here $\sigma$ is the so-called coefficient of deactivation of the earthquake source, "cooling down" after the main shock. For $\sigma = \text{const}$, the general solution of the differential equation (2) coincides with the Omori formula (1) up to notation.

Hirano [9], Utsu and others [10, 11] looked at Omori's law from a different point of view. They decided that the formula (1) is incorrect and offered their own version of the law:

$$n(t)=\frac{k}{(c+t)^p}. \qquad (3)$$

We prefer the law in the form (2). It is formulated using only three symbols ($n$, $t$, $\sigma$) instead of five symbols ($n$, $t$, $k$, $c$, $p$) that are used in formula (3). Moreover, the evolution equation (2) admits natural generalizations [3, 7, 12], which are by no means obvious when formulating the Omori law in the Hirano-Utsu form.

We write the solution of equation (2) in the form closest to the Omori formula (1):

$$n(\tau)=\frac{1}{n_0^{-1}+\tau}. \qquad (4)$$

Here $n_0 = n(0)$. The physical quantity

$$\tau = \int_0^t \sigma(t')dt' \qquad (5)$$

will be called *the proper time* of the source "cooling down" after the main shock of the earthquake. Here we follow the tradition, designating proper time with the letter $\tau$ to distinguish it from world (coordinate) time $t$.. The difference between $\tau$ and $t$ observed by us in the experiment is explained by the non-stationarity of the rock mass in the source after the formation of the main rupture.

To date, successful experience has been gained in using $\tau$ along with $t$ in the study of aftershocks [3–7, 12, 13]. It should be emphasized that earlier the idea of proper time was used implicitly, and only in a relatively narrow area related to the physics of aftershocks. In this paper, we will make an attempt to expand our understanding of the proper time of tectonic processes in the hope that this will be useful in the future. Our attempt is justified by the fact that non-stationarity is an immanent property of the Earth's lithosphere.

## 2. Generalization

As an object, we choose the global activity of strong earthquakes with magnitude $M \geq 7$. By design, we need an "underground clock", the ticking of which marks the passage of proper time. Formula (4) suggests that it is reasonable to try to use relatively weak earthquakes for counting proper time. We will select earthquakes with magnitudes $6 \leq M < 7$ as a trial run.

To illustrate how we can use the proper time, we will do a little test. We presented the sequence of earthquakes as a random process of the Poisson type, i.e. as a chain of instantaneous events separated by some intervals of time. Let us compare the Poisson distribution (see Application)

$$p_k = \frac{\lambda^k}{k!}\exp(-\lambda), \qquad (6)$$

with the distributions for two sequences of strong earthquakes. Let's order one of the sequences according to calendar time, and the second according to proper time.

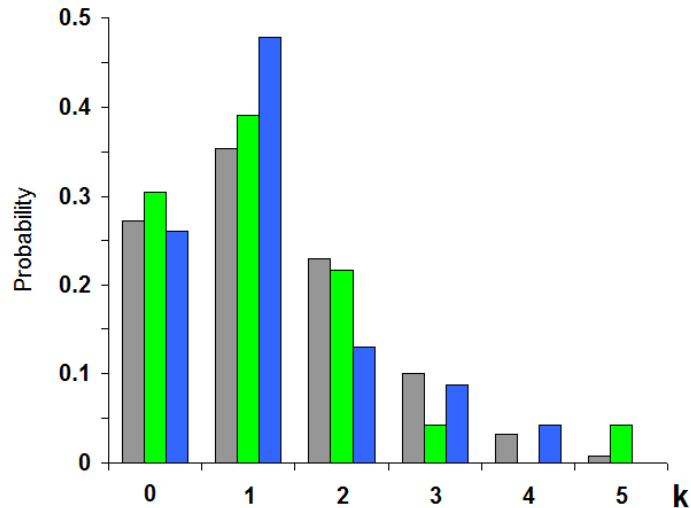

Fig. 1. Distributions of strong earthquakes. The horizontal axis indicates the number of events in the interval of one month of calendar time (blue color) and in the interval of proper time (green color). The gray color represents the Poisson distribution.

The test result is shown in Figure 1. We took data on the time distribution of earthquakes that occurred in 2020 and 2021 from the USGS/NEIC world catalog of earthquakes (https://earthquake.usgs.gov). For 24 months there were 31 strong earthquakes, i.e. on average, $\lambda = 1.3$ events occurred in one month. For 24 months, 255 weaker earthquakes were observed, which we agreed to use for counting prjper time. In Figure 1, the Poisson distribution is colored grey. Blue and green colors show the distributions of events sorted by universal and proper times, respectively. Generally speaking, the distribution of events ordered by proper time is slightly closer to the Poisson distribution than the distribution of events ordered by calendar time. However, the sample size in this illustrative example is too small for us to attach special importance to this. In the next section of the article, we will outline a plan for a more thorough study of the sequence of events, ordered by the proper time.

Let's make a comparative analysis on a more extensive numerical material. First, note that in a Poisson process, events occur uniformly on average. This means that over time the Poisson process, on average, leads to a linear increase in the accumulated number of events (e.g., see Application).

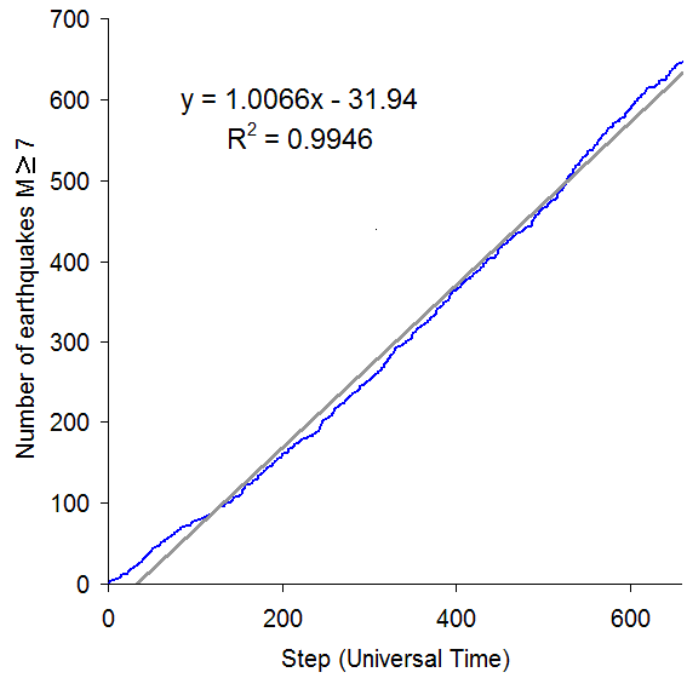

Fig. 2. Dependence of the accumulated number of earthquakes with $M \geq 7$ on universal time. The time step of 26 days is chosen so that the number of points in Figure 2 approximately coincides with the number of points in Figure 3.

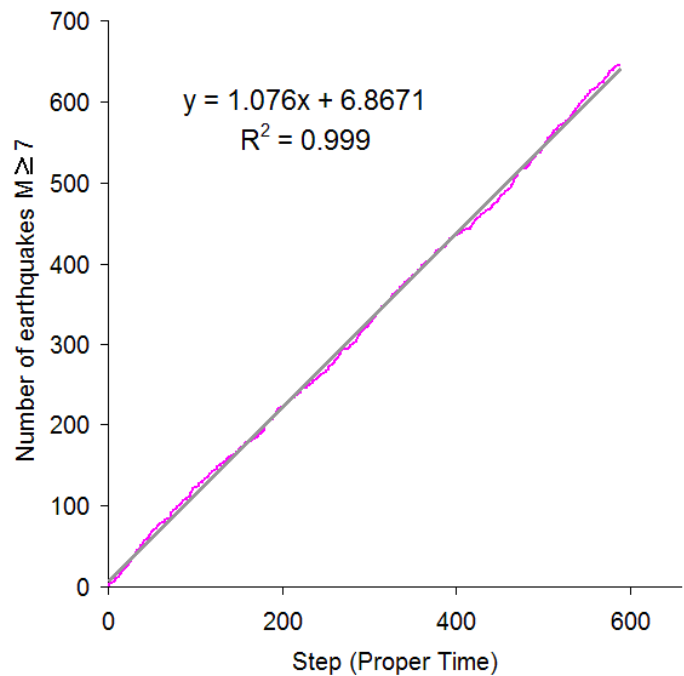

Fig. 3. Dependence of the accumulated number of earthquakes with $M \geq 7$ on proper time. The accumulation step is equal to 10 strokes of the "underground clock".

From 1973 to 2019, 646 earthquakes with magnitudes $M \geq 7$ and 5886 earthquakes with magnitudes $6 \leq M < 7$ occurred on the Earth. The course of accumulation of strong earthquakes is shown in Figures 2 and 3 when the events are ordered according to the universal time and proper time, respectively. We see that in the second case the experimental curve deviates less from a straight line than in the first.

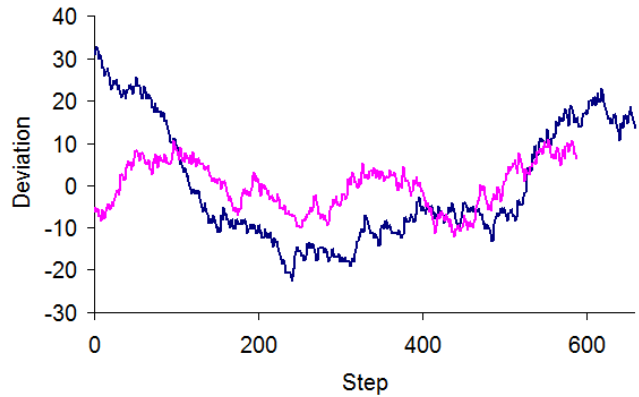

Fig. 4. Deviation of the curves shown in Figures 2 (blue) and 3 (red) from the theoretical straight line.

In order to make this difference more clear, Figure 4 shows the deviations of the real curves from a straight line. It is obvious that the use of proper time corresponds better to the theory of the Poisson process than the use of universal time. This is even more convincingly evidenced by Figure 5, which shows the distributions of the absolute deviations of the experimental curves from the theoretical lines.

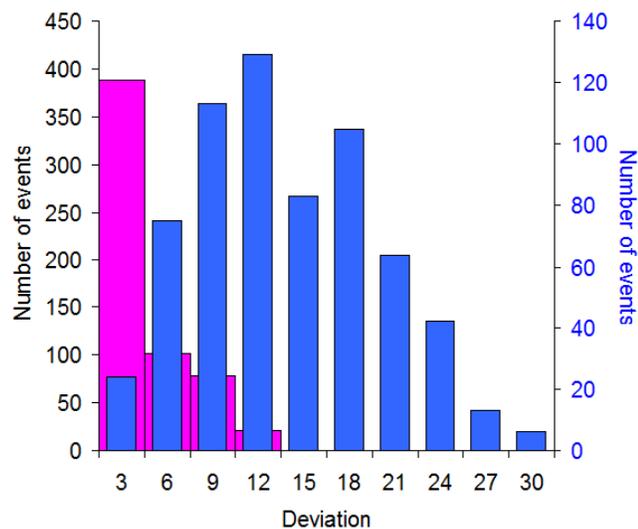

Fig. 5. The distributions of the absolute values of deviation of the curves shown in Figures 2 (blue) and 3 (red) from straight lines.

## 3. Discussion

The geological environment, generally speaking, is non-stationary, its parameters change in time. It is very important that non-stationarity is often hidden from the observer. When studying geodynamic phenomena using world time clocks, the observer detects some correlations, however, in the presence of latent nonstationarity, correlations may turn out to be weak, or not be observed at all when ordering the sequence of phenomena according to world time clocks. In these circumstances, a less perfect clock that keeps the own proper time can be a useful tool for the researcher.

Omori's hyperbolic law in the form (2) allows us to give an illustrative example that explains the above. We measure the flow of proper time by aftershocks. We introduce an auxiliary function $g(t) = 1/n(t)$ and average it over appropriately chosen small time intervals: $g \to \langle g \rangle$. After that, we calculate the deactivation coefficient of the earthquake source using the formula

$$\sigma = \frac{d}{dt}\langle g \rangle. \qquad (7)$$

In the so-called Omori epoch [7], when $\sigma = \text{oconst}$, there is a uniform growth of the parameter $\langle g \rangle$ over time: $\langle g \rangle \propto t$.

We briefly described the formulation and procedure for solving the inverse problem of source physics [2–4, 13]. Experience has shown that the deactivation coefficient, this most important phenomenological parameter of the source, experiences complex variations over time. In other words, the rocks in the source are in a non-stationary state. Note that when setting and solving the inverse problem, the idea of proper time was used implicitly.

Earlier, Hirano [9] already drew attention to the non-stationarity of the geological medium in the cooling chamber, and tried to take into account the non-stationarity by replacing the parameter $p$ in formula (3) with a piecewise-continuous function $p(t)$. However, with such a replacement, formula (3) becomes logically unintelligible. In contrast, Omori's law in the form (2), (4) is mathematically correct, contains the concept of proper time, and allows flexible modeling of the non-stationarity of the medium in the earthquake source.

In an attempt to generalize the idea of proper time, we have chosen a "clock" that, figuratively speaking, is ticking too loudly. In fact, the counting of time by registering earthquakes with magnitudes from the interval $6 \leq M < 7$ is motivated only by the simplicity of processing a relatively small array of numerical information. In the future, we plan to use weaker earthquakes.

The choice of a specific object of study (global activity of strong earthquakes) was also made rather arbitrarily in this work. In the future, it will be interesting to study not only global, but also regional seismicity. The subject of study may be, for example, the cross-correlation of earthquakes in Northern and Southern California. From 1983 to 2007, an anticorrelation was found between fluctuations in the average daily magnitude of earthquakes in these two adjacent regions [14]. We can try to refine this result by using proper time to order earthquakes.

## 4. Conclusion

In this paper, which is purely debatable, we explicitly introduced into geodynamics the concept of proper time, which is different from universal time. Earlier, we and our colleagues have already used the idea of proper time, but implicitly, and only in a relatively narrow area related to the physics of aftershocks. In this paper, we have tried to expand the range of applicability of the concept of proper time, indicated a possible way to measure it, and provided a simple example illustrating the procedure for ordering a sequence of strong earthquakes in proper time. We will continue to study the properties of proper time and plan to submit an extended version of this work to the journal "Geodynamics and Tectonophysics".

*Acknowledgments*. We express our sincere gratitude to B.I. Klain and A.D. Zavyalov for their unfailing support and fruitful discussions. We thank the staff of the US Geological Survey for providing earthquake catalogs.

### Application: Poisson distribution

Gantsevich [15] proposed an interesting derivation of formula (6). Let us briefly describe the course of his reasoning. Let's start from the axiom

$$\sum_{k=0}^{\infty} p_k = 1. \quad (A1)$$

We use the identity

$$1 = \exp(\lambda)\exp(-\lambda). \quad (A2)$$

We represent the exponent in the form of the Taylor series

$$\exp(\lambda) = \sum_{k=0}^{\infty} \frac{\lambda^k}{k!}. \quad (A3)$$

Combining (A1)–(A3), we obtain

$$\sum_{k=0}^{\infty} p_k = \sum_{k=0}^{\infty} \frac{\lambda^k}{k!} \exp(-\lambda), \quad (A4)$$

whence (6) follows.

When $\lambda = at$, the value of $p_k(t)$ is the probability that $k$ events occur during the time $t$. The average number of events is

$$\langle k \rangle = \sum_{k=0}^{\infty} k p_k(t) = at. \tag{A5}$$

When $\lambda = \alpha\tau$, the value $p_k(\tau)$ is the probability that $k$ events occur in time $\tau$. The average number of events is

$$\langle k \rangle = \sum_{k=0}^{\infty} k p_k(\tau) = \alpha\tau. \tag{A6}$$

## References

1. *Guglielmi A.V.* Interpretation of the Omori law // arXiv:1604.07017 [physics.geo-ph] // Izvestiya, Physics of the Solid Earth. 2016. V. 52. P. 785–786.
2. *Guglielmi A.V.* Omori's law: a note on the history of geophysics // Phys. Usp. 2017. V. 60. P. 319–324. DOI: 10.3367/UFNe.2017.01.038039.
3. *Zavyalov A.D., Guglielmi A.V., Zotov O.D.* Three problems in aftershock physics // J. Volcanology and Seismology. 2020. V. 14. No. 5. P. 341–352. DOI: 10.1134/S0742046320050073.
4. *Guglielmi A.V., Klain B.I. Zavyalov A.D., Zotov O.D.* A Phenomenological theory of aftershocks following a large earthquake // J. Volcanology and Seismology. 2021. V. 15. No. 6. P. 373–378.
5. *Guglielmi A., Zotov O.D.* Dependence of the source deactivation factor on the earthquake magnitude // arXiv:2108.02438 [physics.geo-ph]. 2021.
6. *Zotov O.D., Guglielmi A.V.* Mirror triad of tectonic earthquakes // arXiv:2109.05015 [physics.geo-ph].
7. *Guglielmi A.V., Zotov O.D., Zavyalov A.D., Klain B.I.* On the fundamental laws of earthquake physics // J. Volcanology and Seismology, 2022, Vol. 16, No. 2, pp. 143–149. DOI: 10.1134/S0742046322020026
8. *Omori F.* On the aftershocks of earthquake // J. Coll. Sci. Imp. Univ. Tokyo. 1894. V. 7. P. 111–200.
9. *Hirano R.* Investigation of aftershocks of the great Kanto earthquake at Kumagaya // Kishoshushi. 1924. Ser. 2. № 2. P. 77.
10. *Utsu, T.*, A statistical study on the occurrence of aftershocks // Geophys. Mag. 1961. V. 30. P. 521–605.
11. *Utsu T., Ogata Y., Matsu'ura R.S.* The centenary of the Omori formula for a decay law of aftershock activity // J. Phys. Earth. 1995. V. 43. № 1. P. 1–33.
12. *Guglielmi A.V., Klain B.I.* The phenomenology of aftershocks // arXiv:2009.10999.
13. *Guglielmi A.V., Zotov O.D., Zavyalov A.D.* A project for an Atlas of aftershocks following large earthquakes // J. Volcanology and Seismology. 2019. V. 13. No. 6. P. 415–419. DOI: 10.1134/S0742046319060034 // Atlas of aftershock sequences of strong earthquakes. // In: Yanovskaya T., Kosterov A., Bobrov N., Divin A., Saraev A., Zolotova

N. (eds). Problems of Geocosmos-2018. Springer Proceedings in Earth and Environmental Sciences. Springer, Cham. 2020. P. 193–198. DOI: 10.1007/978-3-030-21788-4_15.
14. *Zotov O.D., Zavyalov A.D., Guglielmi A.V., Klain B.I.* Features of the dynamics of seismic activity in Northern and Southern California // 6th International Conference "Trigger Effects in Geosystems". Abstracts of reports. June 21-24, 2022. IDG RAS. Moscow (https://conf2022.idg.ras.ru/schedule.php).
15. *Gantsevich S.V.* Diagrams, resolvent, probability. IV. Poisson process and path integrals // FTI RAS, St. Petersburg. 02/18/2013. 189.25 Kb. (In Russian.)